\documentclass[11pt]{elsart}
\usepackage{color,cite}
\usepackage{amssymb,epsfig}%%,subfigure} -- this is really silly

\def\ifm#1{\relax\ifmmode#1\else$#1$\fi}

\def\to{\ifm{\rightarrow}}
\def\ab{\ifm{\sim}}  
\def\x{\ifm{\times}}
\def\f{\ifm{\phi}}  
\def\DAF{DA\char8NE}   
\def\ff{\f--factory}
\def\ppc{\ifm{\pi^+\pi^-}}  
\def\ppo{\ifm{\pi^0\pi^0}}  
\def\pppco{\ifm{\pi^+\pi^-\pi^0}}  
  
\def\pppo{\po\po\po}  
\def\up#1{\ifm{^{#1}}}  

\def\ks{\ifm{K_S}} 
\def\kl{\ifm{K_L}}

\def\po{\ifm{\pi^0}}

\let\cal=\mathcal   
\def\ORD#1!{\ifm{{\cal O}\hbox{(#1)}}}
\def\pt#1,#2,{\ifm{#1\x10^{#2}}} 

\def\bye{\end{document}}

\newcount\figurecount    
\def\be{\begin{equation}}
\def\ee{\end{equation}}  
\def\bea{\begin{eqnarray}}
\def\eea{\end{eqnarray}}
\def\bc{\begin{center}}
\def\ec{\end{center}}  
\def\ben{\begin{enumerate}}
\def\een{\end{enumerate}}

\def\sms{\kern-1mm}

\def\figbox#1;#2;{\parbox{#2cm}{%
\vglue1mm\epsfig{file=#1.eps,width=#2cm}\vglue1mm}}

\makeatletter
\newdimen\z@ \z@=0pt % can be used both for 0pt and 0
\newskip\z@skip \z@skip=0pt plus0pt minus0pt
\def\m@th{\mathsurround=\z@}
\def\ialign{\everycr{}\tabskip\z@skip\halign} % initialized \halign
\def\eqalign#1{\null\,\vcenter{\openup\jot\m@th
  \ialign{\strut\hfil$\displaystyle{##}$&$\displaystyle{{}##}$\hfil
      \crcr#1\crcr}}\,}
\makeatother

\newcommand{\aff}[2]{Dipartimento di Fisica dell'Universit\`a di #1 e Sezione INFN, #2, Italy.}
\newcommand{\affd}[1]{Dipartimento di Fisica dell'Universit\`a e Sezione INFN, #1, Italy.}

\begin{document}
\begin{frontmatter}


\title{Measurement of the $K_L$ meson lifetime with the KLOE detector}
\collab{The KLOE Collaboration}

\author[Na]{F.~Ambrosino},
\author[Frascati]{A.~Antonelli},
\author[Frascati]{M.~Antonelli},
\author[Roma3]{C.~Bacci},
\author[Frascati]{P.~Beltrame},
\author[Frascati]{G.~Bencivenni},
\author[Frascati]{S.~Bertolucci},
\author[Roma1]{C.~Bini}
\author[Frascati]{C.~Bloise},
\author[Roma1]{V.~Bocci},
\author[Frascati]{F.~Bossi},
\author[Frascati,Virginia]{D.~Bowring},
\author[Roma3]{P.~Branchini},
\author[Roma1]{R.~Caloi},
\author[Frascati]{P.~Campana},
\author[Frascati]{G.~Capon},
\author[Na]{T.~Capussela},
\author[Roma3]{F.~Ceradini},
\author[Frascati]{S.~Chi},
\author[Na]{G.~Chiefari},
\author[Frascati]{P.~Ciambrone},
\author[Virginia]{S.~Conetti},
\author[Roma1]{E.~De~Lucia},
\author[Roma1]{A.~De~Santis},
\author[Frascati]{P.~De~Simone},
\author[Roma1]{G.~De~Zorzi},
\author[Frascati]{S.~Dell'Agnello},
\author[Karlsruhe]{A.~Denig},
\author[Roma1]{A.~Di~Domenico},
\author[Na]{C.~Di~Donato},
\author[Pisa]{S.~Di~Falco},
\author[Roma3]{B.~Di~Micco},
\author[Na]{A.~Doria},
\author[Frascati]{M.~Dreucci},
\author[Frascati]{G.~Felici},
\author[Roma3]{A.~Ferrari},
\author[Frascati]{M.~L.~Ferrer},
\author[Frascati]{G.~Finocchiaro},
\author[Frascati]{C.~Forti},
\author[Roma1]{P.~Franzini},
\author[Frascati]{C.~Gatti},
\author[Roma1]{P.~Gauzzi},
\author[Frascati]{S.~Giovannella},
\author[Lecce]{E.~Gorini},
\author[Roma3]{E.~Graziani},
\author[Pisa]{M.~Incagli},
\author[Karlsruhe]{W.~Kluge},
\author[Moscow]{V.~Kulikov},
\author[Roma1]{F.~Lacava},
\author[Frascati]{G.~Lanfranchi\corauthref{cor1}}
\author[Frascati,StonyBrook]{J.~Lee-Franzini},
\author[Roma1]{D.~Leone},
\author[Frascati]{M.~Martini},
\author[Na]{P.~Massarotti},
\author[Frascati]{W.~Mei},
\author[Na]{S.~Meola},
\author[Frascati]{S.~Miscetti},
\author[Frascati]{M.~Moulson},
\author[Karlsruhe]{S.~M\"uller},
\author[Frascati]{F.~Murtas},
\author[Na]{M.~Napolitano},
\author[Roma3]{F.~Nguyen},
\author[Frascati]{M.~Palutan},
\author[Roma1]{E.~Pasqualucci},
\author[Roma3]{A.~Passeri},
\author[Frascati,Energ]{V.~Patera},
\author[Na]{F.~Perfetto},
\author[Roma1]{L.~Pontecorvo},
\author[Lecce]{M.~Primavera},
\author[Frascati]{P.~Santangelo},
\author[Roma2]{E.~Santovetti},
\author[Na]{G.~Saracino},
\author[Frascati]{B.~Sciascia},
\author[Frascati,Energ]{A.~Sciubba},
\author[Pisa]{F.~Scuri},
\author[Frascati]{I.~Sfiligoi},
\author[Frascati]{T.~Spadaro},
\author[Roma1]{M.~Testa},
\author[Roma3]{L.~Tortora},
\author[Frascati]{P.~Valente},
\author[Karlsruhe]{B.~Valeriani},
\author[Pisa]{G.~Venanzoni},
\author[Roma1]{S.~Veneziano},
\author[Lecce]{A.~Ventura},
\author[Roma3]{R.~Versaci}
\author[Frascati,Beijing]{G.~Xu}
\address[Frascati]{Laboratori Nazionali di Frascati dell'INFN, 
Frascati, Italy.}
\address[Karlsruhe]{Institut f\"ur Experimentelle Kernphysik, 
Universit\"at Karlsruhe, Germany.}
\address[Lecce]{\affd{Lecce}}
\address[Na]{Dipartimento di Scienze Fisiche dell'Universit\`a 
``Federico II'' e Sezione INFN,
Napoli, Italy}
\address[Pisa]{\affd{Pisa}}
\address[Energ]{Dipartimento di Energetica dell'Universit\`a 
``La Sapienza'', Roma, Italy.}
\address[Roma1]{\aff{Roma, ``La Sapienza''}{Roma}}
\address[Roma2]{\aff{Roma 2, ``Tor Vergata''}{Roma}}
\address[Roma3]{\aff{Roma 3}{Roma}}
\address[StonyBrook]{Physics Department, State University of New 
York at Stony Brook, USA.}
\address[Virginia]{Physics Department, University of Virginia, USA.}
\address[Beijing]{Permanent address: Institute of High Energy 
Physics, CAS,  Beijing, China.}
\address[Moscow]{Permanent address: Institute for Theoretical 
and Experimental Physics, Moscow, Russia.}
\address[Tbilisi]{Permanent address: High Energy Physics Institute, Tbilisi
  State University, Tbilisi, Georgia.}\vglue2mm
\corauth[cor1]{cor1}{\vbox{\raggedright\small $^1$ Corresponding author: Gaia Lanfranchi, INFN - LNF, CP 13, 00044 Frascati (Roma), 
Italy; e-mail: gaia.lanfranchi@lnf.infn.it}}
\begin{abstract}
\noindent 
We present a measurement of the $\kl$ lifetime using 
the KLOE detector. 
From a sample of $\sim 4 \times 10^8$ $K_S K_L$ pairs
following the reaction $e^+ e^- \to \phi \to K_S K_L$ 
we select $\sim 15 \times 10^{6} $
$\kl \to \pppo$ decays tagged by $\ks \to \ppc$ events. 
From a fit of the proper time distribution
we find $\tau_L = (50.92 \pm 0.17_{\rm stat} \pm 0.25_{\rm syst})$ ns.
This is the most precise measurement of the $K_L$ lifetime performed to date.

\par\noindent PACS:
\par\noindent keywords:

\end{abstract}
\end{frontmatter}
\overfullrule=10pt
%---------------------------------------------
\section{Introduction}
%---------------------------------------------
The $\kl$ lifetime is necessary to determine its semileptonic partial widths from the branching ratios (BR). The partial widths can be used to extract the CKM matrix element $|V_{us}|$.
Present knowledge of $\tau(\kl)$ comes from a single measurement performed more than 30 years ago \cite{vosburgh:tkl} and its error dominates the uncertainty in the partial $K_L$ decay rates. At \DAF\ , the Frascati \ff\ ,
nearly monochromatic $K_L$-mesons are produced with $p$\ab110 MeV/c corresponding to a mean path of 340 cm. The KLOE detector is large enough, $r$=200 cm, so that \ab50\% of the \kl\ decay inside it.
The statistical error on the lifetime depends strongly 
on the time interval covered in the measurement \cite{K-note}:

\begin{equation} 
{\delta \tau \over \tau } = {1 \over \sqrt{N}} \times 
\left [ {-1 + e^{3T} + (e^T-e^{2T})(3+T^2) \over (-1+e^T)^3 }\right]^{-0.5}
\label{eq:paolo} 
\end{equation}

where $T=\Delta t/\tau$ is the time interval observed, in $K_L$-lifetime units.
With $T$\ab0.4 and $N$\ab\pt9,6, we can reach an accuracy of \ab0.3\%.

 We have measured the \kl\ lifetime using the decay \kl\ $\to$ \pppo\ tagged by 
\ks\to\ppc\ events. This choice maximizes the number of usable events and minimize the disturbance of the \kl\ decay on the detection of the tagging \ks\ decay and therefore the systematic uncertainty.

%---------------------------------------------
\section{Experimental setup}
%---------------------------------------------

In DA$\Phi$NE electrons and positrons collide with an angle of 25 mrad and a center of mass (CM) energy $W=M(\phi)$. 
$\phi$-mesons are produced with a cross section of \ab3 $\mu$b and a transverse momentum of $\sim$ 12.5 MeV/c toward the center of the collider rings.
The energy $W$, the position of the beam crossing point ($x,y,z$) and the $\phi$
momentum are determined from Bhabha scattering events. In a typical run of integrated luminosity $\int {\cal L}$dt\ab100 nb$^{-1}$, lasting about 30 minutes, the corresponding errors are: $\delta W$=40 keV, $\delta p_{\phi}$=30
keV/c, $\delta x = 30\ \mu$m, and $\delta y = 30\ \mu$m.

The detector consists of a large cylindrical drift chamber, DC
\cite{K-DC}, whose axis, defined as the $z$-axis, coincides with the bisectrix of the two beams. The DC is surrounded by a lead-scintillating fiber sampling
calorimeter, EMC \cite{K-EMC}. The DC and EMC are immersed in a solenoidal magnetic
field of 0.52 T with the axis parallel to the beams' bisectrix.
The DC tracking volume extends from 28.5 to 190.5 cm
in radius and is 340 cm long. The transverse momentum resolution is $\delta p_\bot/p_\bot$\ab0.4\%. Vertices are reconstructed with a resolution of
\ab3 mm. 
The calorimeter is divided into a barrel and two endcaps
and covers 98$\%$ of the solid angle. 
Photon energies and
arrival times are measured with resolutions $\sigma_{E}/E =
0.057/\sqrt{E\ ({\rm GeV})}$  and $\sigma_{t}=54\ {\rm ps}/\sqrt{E\ ({\rm GeV})}\oplus50\ {\rm ps}$  respectively. 
Photon entry points are determined with an accuracy $\sigma_z\ab1\ {\rm cm}/\sqrt{E\ ({\rm GeV})}$ along the fibers and $\sigma_\bot$\ab1 cm in the transverse direction.
A photon is defined as an EMC cluster of energy deposits not associated to a track. We require that the distance between the cluster centroid and the entry point of the nearest extrapolated track be greater than 3$\sigma$, $\sigma = \sigma_z \oplus  \sigma_\bot$.

The trigger \cite{K-TRIGGER} uses information from the calorimeter and chamber. The EMC trigger requires two local energy deposits above threshold ($E > 50$ MeV in the barrel, $E > 150$ MeV in the endcaps). 
Rejection of cosmic-ray events is also performed at trigger level, checking for the presence of two energy deposits above 30 MeV in the outermost calorimeter planes.
The DC trigger is based on the multiplicity and topology of the hits in the drift cells. 
The trigger has a large time spread  with respect to the beam crossing
time. It is therefore re-synchronized with the machine radio frequency divided by four, $T_{\rm sync}$ = 10.85 ns, with an accuracy of 50 ps.
During the 2001-2002 data taking the bunch crossing period at \DAF\ was $T$=5.43 ns. 
The correct collision time, $T_0$, of the event is determined off-line during  event reconstruction \cite{K-OFFLINE}.

%---------------------------------------------
\section{Data analysis}
%---------------------------------------------

$\phi$-mesons decay into \ks-\kl\ pairs \ab34\% of the time. Production of a $K_{L}$ is tagged
by the observation of \ks\to\ppc\ decay. The \kl\to\pppo\ decay vertex is reconstructed along the direction opposite to that of the \ks\ in the $\phi$ rest frame. The chamber alone measures the \ks\to\ppc\ decay and therefore the direction of the $K_{L}$. The $K_{L}$ decay vertex and 
the photon energies are obtained from EMC information.
The data sample, collected during 2001 and 2002, corresponds to an integrated luminosity of \ab400 pb\up{-1}. Some \pt1.2,9, $\phi$-mesons were produced. Additional details can be found in reference \cite{K-note}.
$K_{S} \to \pi^{+} \pi^{-}$ decay events must satisfy the following requirements:
\begin{enumerate}
\item There must be two tracks with opposite charge, forming a vertex V in a cylinder with $r_{\rm V}< 10$ cm, $|z_{\rm V}| < 20$ cm. No other tracks should be connected to the vertex.
\item The $K_S$ momentum in the $\phi$ rest system, must satisfy $100< p_{K_S}<120$ MeV/c. The \ppc\ invariant mass $M(\pi\pi)$ must satisfy 492$<M(\pi\pi)<$503 MeV/c$^2$.
\end{enumerate}

The efficiency for finding $K_S \to \pi^+ \pi^-$ events is $\epsilon \sim 68 \%$.
The position of the $\phi$ production point, ${\bf x_{\phi}}$, is determined as the point of closest approach of the $K_S$ momentum, propagated backwards
from the $K_S$ vertex, to the beam line.
The $K_S \to \pi^+ \pi^-$ decay provides an almost unbiased tag for the $K_{L}$ when it
decays into neutral particles and a good measurement of the $K_{L}$ momentum, $\vec p_{\kl}= \vec p_\phi-\vec p_{\ks}$. 
The accuracy in the determination of the $K_{L}$ direction is obtained from \kl\to\ppc\po\ events, measuring the angle between $\vec p_{\kl}$ and the line joining the $\phi$ production point and the \pppco\ decay vertex.
We find $\sigma_\phi = 1.5^\circ$, $\sigma_\theta = 1.8^\circ$.

\begin{figure}[htb]
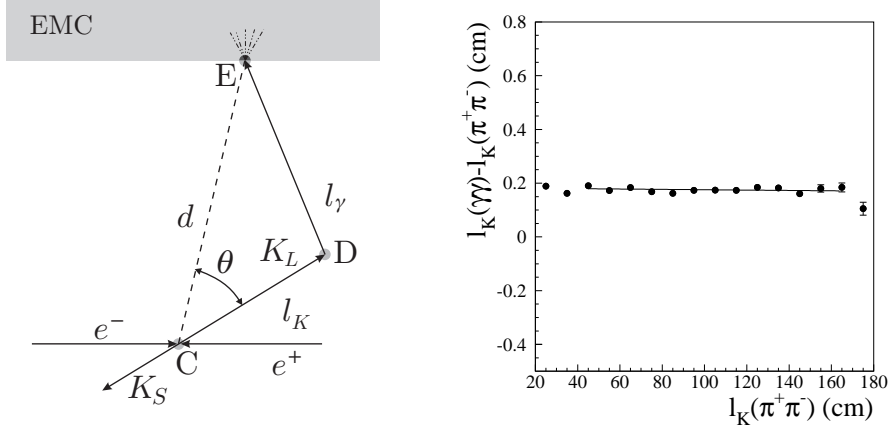

\centerline{\figbox l-k;5;\kern10mm\figbox timescale_all_new;6;}
\caption{Left. The CDE triangle. Right. Distribution of the difference $l_K(\po)\break-l_K(\ppc)$ for $K_L \to \pi^+ \pi^- \pi^0$ events as a function of $l_K(\ppc)$. See text.}
 \label{FIG1}
\end{figure}

The position of the $K_L$ vertex for \kl\to\pppo\ decays is obtained from the photon arrival times at the EMC.
Each photon defines a triangle CDE, see figure \ref{FIG1} (left), where $l_K$ is the \kl\ path length, $l_{\gamma}$ is the distance from the \kl\ decay point D to the entry point E and $d$ is 
the distance from the cluster to the collision point C.
From the known positions of C and E, the $\widehat{ECD}=\theta$ angle and the time spent by the kaon and the photon to cover the path CDE we find the length of CD.
There are two solutions. One has D along the \ks\ path and is rejected.
The position D of the \kl\ decay vertex is obtained from the energy 
weighted average of the two closest $l_{K,i}$, 
$l_{K} = \sum(l_{K,i}\times E_i)/\sum E_i$  where $i$ 
is the photon index. Finally we require at least one third photon with $|l_{K,3}-l_K|<5\x\sigma(l_K)$.

The accuracy of the $l_K$ determination is checked by
comparing the \kl\ path measured by timing with the
calorimeter and, with a much better accuracy, by tracking with the DC, for \kl\to\ppc\po\ decays.
The path length from the calorimeter timing has on average a constant offset of 2 mm with respect to the value obtained with the DC, figure \ref{FIG1}, right.
The determination of $l_K$ depends crucially on the correct identification of the collision time $T_0$. Using again \kl\to\ppc\po\ events we have verified that $T_0$ is incorrect less than 0.1\% of the time.

\begin{figure}[htb]
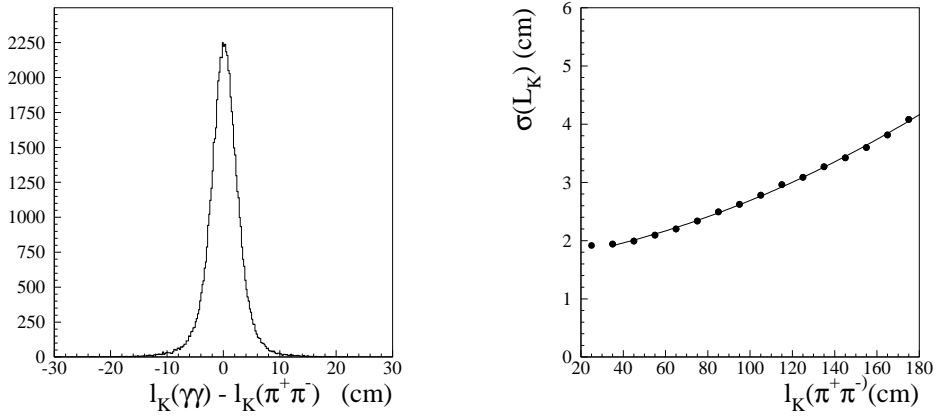

\centerline{\figbox plot_res_null;6;\kern10mm\figbox sigma_vs_lk;6;}
\caption{Left. Distribution of the difference $l_K(\po)-l_K(\ppc)$ 
for $K_L \to \pi^+ \pi^- \pi^0$ events. Right. The $\sigma$ obtained with 
a single-gaussian fit as a function of $l_K(\ppc)$. See text.}
 \label{fig:res}
\end{figure}

The resolution $\sigma(l_K)$ is determined from $K_{L} \to \pi^{+} \pi^{-} \pi^{0}$ 
events by comparing $l_K(\po)$ and $l_K(\pi^+ \pi^-)$,
where the former is the weighted average obtained from 
the two photons from $\pi^0$ and $l_K(\pi^+ \pi^-)$ 
is the distance between the vertex of the two charged decay pions and the $\phi$ production point.
%-----new
An example of the  $l_K(\po) - l_K(\pi^+ \pi^-)$ distribution is shown in figure \ref{fig:res}, left.
It has been fitted both with a single and a double gaussian \cite{K-note}.
In the case of the double-gaussian fit the relative weights of the two components are free
parameters of the fit.
The single-gaussian fit gives an average resolution of $\sim 2.5$ cm.
In the double-gaussian fit, the bulk of the distribution
($\sim 82 \%$ of the total) has  $\sigma_1 \sim 2.1$ cm,
while the broader part ($\sim 18 \%$) is well described 
by a gaussian with $\sigma_2 \sim 5.4$ cm.
The behaviour of the resolution has been studied as a function of $l_K(\ppc)$:
we find a quadratic dependence on $l_K$ in both cases (single and double-gaussian fit). 
For the single-gaussian fit we have 
$\sigma(l_K)=1.65+\pt0.59,-2,\x l_K +\pt0.45,-4,\x l_K^2$ cm (figure \ref{fig:res}, right). 
In the double-gaussian fit the relative weights of the two components
change as a function of $l_K(\ppc)$.
Since for each point the population weighted average  of $\sigma_1$ and $\sigma_2$ agrees at the 10$\%$ level
with the $\sigma$ of a single-gaussian fit, we use the latter as estimate of the vertex resolution.
The effects of the tails on the fit value are discussed in Section 6.

%-----new

\begin{figure}[hb]
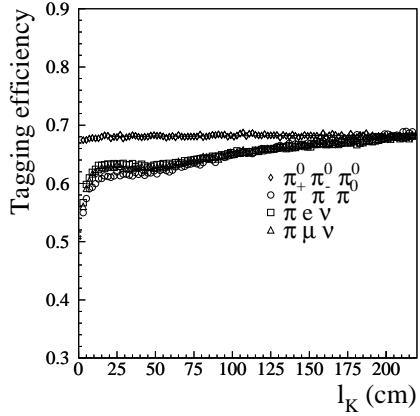

\centering 
\figbox tag1_50_new;6;
\caption{Tagging efficiency as a function of $l_K$ for the 
main \kl\ decay modes.}
 \label{fig:tag}
\end{figure}

The tagging efficiency has been evaluated by MC as a function
of $l_K$ for the dominant \kl\ decay channels. 
The results are shown in figure \ref{fig:tag}.
The difference in tagging efficiency among the $K_L$ decay modes
is mainly due to the dependence of the trigger efficiency.
Only the  calorimeter trigger \cite{K-TRIGGER}
is used for the present analysis.
The trigger efficiency is, on average, \ab100\% for $\kl \to \pppo$ and
between $ 85-95 \% $ for charged  $\kl$  decays.
The trigger efficiency also depends on the position of the $\kl$  decay vertex.
Another contribution is the dependence
of the reconstruction efficiency for the pion tracks from $\ks \to \ppc$ on the
presence of other tracks in the drift chamber.
This contribution depends on the position of 
the $\kl$ decay point and affects mainly events 
with $\kl \to$ charged particles near the $\phi$ production point.

The tagging efficiency for 
the $\kl \to \pppo$ channel has a small linear dependence on $l_K$, 
with a slope of 
$b = (1.2 \pm 0.2) \times 10^{-5}$/cm,
and a constant $a= (68.04 \pm 0.01) \%$.

%-------------------------------------------------
\section{$K_L \to \pi^0 \pi^0 \pi^0$ acceptance}
%-------------------------------------------------
The \kl\to\pppo\ decay has a relatively large BR, \ab21\%, and has very low background. 
\kl\to\pppo\ events are accepted if at least three calorimeter clusters are found satisfying: \begin{enumerate}
\item Energy larger than 20 MeV.
\item Distance from any other cluster larger than 50 cm.
\item No association to a chamber track.
\item $|l_{K,i}-l_K|<5\times\sigma(l_K)$, where $l_K$ is the
 energy weighted average of the two values of $l_{K,i}$ nearest together.
\end{enumerate}
For the \kl\ lifetime measurement, we retain events with 40 $<l_K<$ 165 cm and a polar angle $\theta$ in the interval $\{40^{\circ},140^{\circ}\}$. 
These two conditions define the fiducial volume (FV).
The main sources of event losses are: 1) geometrical acceptance;
2) cluster energy threshold; 3) merging of clusters; 4) accidental
association to a charged track; 5) Dalitz decay of one or more
$\pi^{0}$'s. The effect of these inefficiencies is to modify the
relative population for events with 3, 4, 5, 6, 7 and $\ge 8$,
clusters with a loss of global efficiency of $\sim 0.8\%$.

\begin{figure}[hbt]
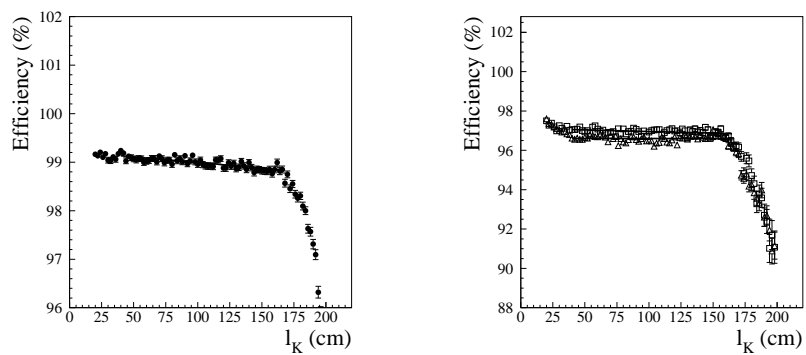

\centerline{\figbox eff3p0_20mev_new;5;\kern10mm\figbox effppp_20mev_new;5;}
\caption{ Left:
   Vertex reconstruction efficiency as a function of the 
   decay path length for  $K_L \to \pi^0 \pi^0 \pi^0$ Monte Carlo events.
    Right: the same for $K_L \to \pi^+ \pi^- \pi^0$ data (triangles) and
   Monte Carlo (squares) events. }
 \label{fig:eff_vtx}
\end{figure}

Monte Carlo (MC) simulations, 
based on the KLOE standard MC \cite{K-OFFLINE} show 
that event acceptance with the above selection 
has a linear dependence on $l_K$, 
$\epsilon(l_K)=(0.9921 \pm 0.002)-\pt(1.9 \pm 0.2),-5,\x l_K$ with $l_K$ in cm,
(figure \ref{fig:eff_vtx}, left)
mainly due to the vertex reconstruction efficiency.
This has also been checked using $\kl \to \pppco$ events both from data and MC.
We find the same linear dependence, with compatible slopes within their statistical uncertainties, figure \ref{fig:eff_vtx}, right.

A comparison between data and MC of the photon multiplicity and 
total energy distributions for $\kl \to \pppo$ decays shows that 
only events with three and four clusters contain some background. 
Background, mostly to the three cluster events, is due to $K_L \to \pi^+ \pi^- \pi^0$ 
decays where one or two charged pions produce a cluster not associated 
to a track and neither track is associated to the $K_L$ vertex. 
Other sources of background are  $K_L \to
\pi^0 \pi^0$ decays (possibly in coincidence with machine background showering close to the collision point generating soft neutral particles) and $ K_S \to \pi^0 \pi^0$ 
following $K_L \to K_S$ regeneration in the DC material.
The $K_L \to \pi^+ \pi^- \pi^0$ background and the other backgrounds in the 
three cluster sample are
strongly reduced by requiring at least one cluster in the barrel with $E\ge$50 MeV and no tracks approaching the $K_L$ line of flight by less than 20 cm. 
The efficiency vs $l_K$ for the three cluster sample 
has been found by MC. It is almost flat with an average \ab55\% inside the FV.

\begin{figure}[htb]
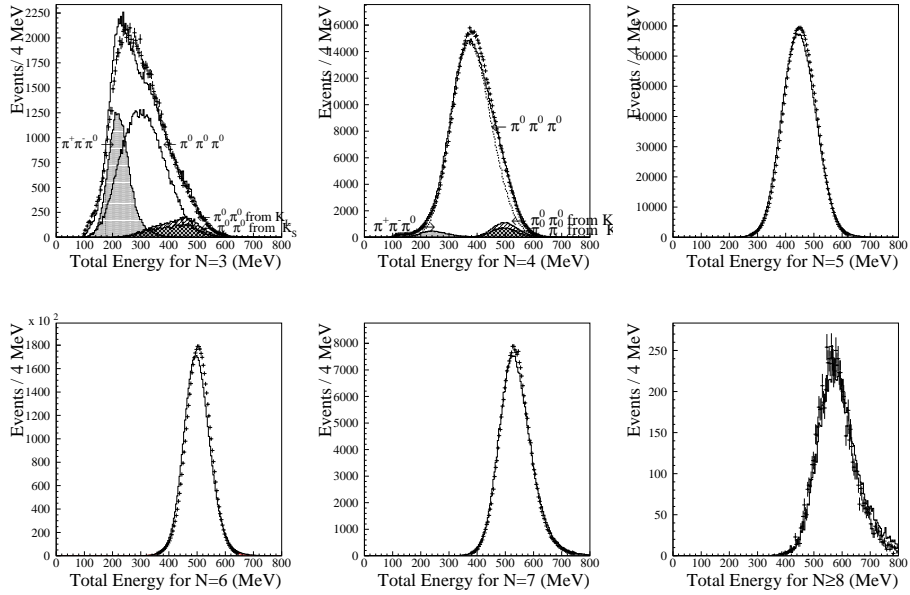

\centerline{\figbox etot3_40-165_new;4;\kern1mm\figbox etot4_40-165_new;4;\kern1mm\figbox etot5_40-165_new;4;}
\centerline{\figbox etot6_40-165_new;4;\kern1mm\figbox etot7_40-165_new;4;\kern1mm\figbox etot8_40-165_new;4;}
\caption{$K_L \to \pi^0 \pi^0 \pi^0$ selection:
    distribution of the total energy for events with
    3, 4, 5, 6, 7 and $\geq$ 8 photon clusters. Dots are data,
    solid histogram is Monte Carlo simulation for
    $K_L \to$ all channels. Monte Carlo histograms are normalized to
    the same number of entries for data.}
    \label{fig:etot}
\end{figure}

In Table \ref{tab:back2} we show the fractions of the 
background components before and after the background cuts.
The background contamination is reduced from $\sim 4.9 \%$ 
to $\sim 1.3 \%$ with an efficiency on the signal of $\sim 99.6 \%$.
The distributions of the total photon energy for events with $3,4,5,6,7,\ge 8$ photons
are shown in figure~\ref{fig:etot}. For three and four photon-cluster 
samples the different contributions from the residual background components
are also shown.

The fractions of events with $N=3,4,5,6,7,\geq 8$ in the FV are given in Table \ref{tab:nclu}, together with MC results.
The few percent differences between data and MC are mostly due 
to a higher proportion of split clusters in the MC than in the data.

\def\vvv{\vphantom{\vrule height4mm depth0pt}}
\begin{table}[ht] 
\bc\vglue3mm
\renewcommand{\arraystretch}{1.1} % enlarge line spacing
\begin{tabular}{|l|r|c|r|c|} \hline
 \vvv &\multicolumn{2}{c|}{$N_\gamma\ge$3 before cuts}&\multicolumn{2}{c|}{$N_\gamma\ge$3 after cuts}     \\ \hline
 \vvv$K_L$ channel         &  Events     & B/(S+B)      & Events     & B/(S+B)\\ \hline
 \vvv signal+backgrounds    &  10,536,674 &              & 10,114,899    &            \\
 all backgrounds       &   518,520   & 4.92 $\%$    &  133,535     & 1.32 $\%$  \\
 $ \kl \to \pppco$     &  325,076    & 3.08 $\%$    &    44,917    & 0.44 $\%$  \\
 $ \kl\to\pi\mu\nu$    &    28,917   & 0.28 $\%$    &    3,583    & 0.03 $\%$  \\
 $ \kl \to\pi e\nu$    &    49,140   & 0.47 $\%$    &    6,062    & 0.06 $\%$  \\ 
 $ \kl \to \ppo$       &   43,436    & 0.41 $\%$    &    42,313   & 0.42 $\%$  \\ 
 $ \kl\to\ks\to\ppo$   &   30,273    & 0.29 $\%$    &    28,166    & 0.28 $\%$  \\
 $ \kl \to other$      &   41,298    & 0.39  $\%$   &    8440    & 0.08 $\%$  \\  \hline\end{tabular}\ec\vglue3mm
\caption{Event types in the fiducial volume from Monte Carlo 
before and after background cuts. The
 background contamination $B/(S+B)$ is also shown.
}\label{tab:back2}
\end{table}

\begin{table}[htb]
\vglue3mm\bc
\renewcommand{\arraystretch}{1.1} % enlarge line spacing
\begin{tabular}{|c|c|c|} \hline
\vvv Number of clusters   &  Data                & Monte Carlo        \\ \hline
\vvv3                 &  1.163 $\pm$ 0.004 $\%$    & 0.980$\pm$ 0.003  \%  \\
   4                  &  7.64  $\pm$ 0.01  $\%$    & 7.01 $\pm$ 0.01   \%  \\
   5                  & 30.22 $\pm$ 0.02  $\%$     & 28.65$\pm$ 0.02   \% \\ 
   6                  & 57.77 $\pm$ 0.03  $\%$     & 60.12$\pm$ 0.03   \%  \\  
   7                  & 3.091 $\pm$ 0.006 $\%$     & 3.074$\pm$ 0.001  \%  \\ 
 $\ge 8$              & 0.106 $\pm$ 0.001 $\%$     & 0.151$\pm$ 0.001  \% \\ \hline
\end{tabular}\ec\vglue3mm
\caption{ \small Fraction of events with 3,4,5,6,7 and $\geq 8$ neutral
 clusters connected to the $K_{L}$ decay vertex in data and Monte Carlo.}
\label{tab:nclu}
\end{table}
%\renewcommand\baselinestretch{1}
%--------------------------------------------------------------------------
\section{Fit of the proper time distribution}
%--------------------------------------------------------------------------

\begin{figure}[ht]
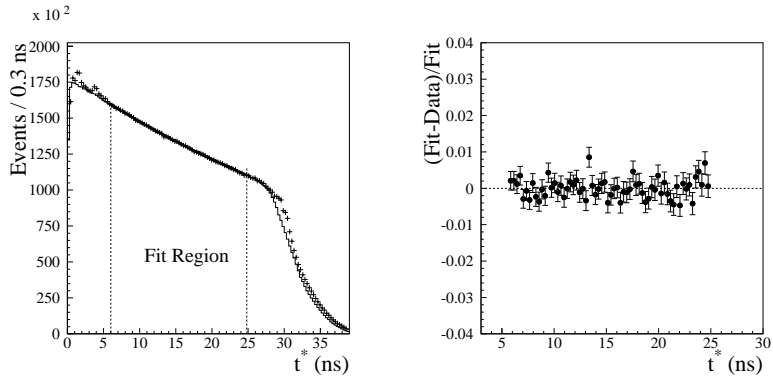

\centerline{\figbox tkl_new;5;\kern5mm\figbox tkl_residui_new;5;}
\caption{ Fit of the proper time distribution (left) and 
 residuals of the fit (right). 
 Crosses are data and solid histogram is Monte Carlo. The fit is shown as the thick solid line.}
 \label{fig:tkl}
\end{figure}

The \kl\ proper time, $t^*$, is obtained event by event dividing 
the decay length $l_K$ by $\beta \gamma$ of the $\kl$  in the laboratory,
$t^* = l_K/(\beta\gamma c)$.
 In figure \ref{fig:tkl} we show the $t^*$
 distribution obtained with $ \sim 14.7 \times 10^6$ tagged $\kl \to \pppo$
 events. 
The residual $\sim 1.3 \%$ background is subtracted using MC results.
The variation of the vertex reconstruction efficiency 
 as a function of the decay length is taken into account by correcting 
 bin by bin the $t^*$ distribution with product of 
the MC $\kl \to 3 \pi^0$ efficiency 
(figure \ref{fig:eff_vtx}, left) and the  data/MC efficiency ratios 
 for $\kl \to \pppco$ (figure \ref{fig:eff_vtx}, right). 
The statistical uncertainty ($\sim 0.1\%$) of the efficiency 
estimate is included in the error. 

%----------- new
Both the background subtraction and the 
vertex reconstruction efficiency correction
affect the number of events per bin at the $\sim 1\%$ level 
and the combined effect of both corrections leaves 
the effective statistics essentially unchanged. 
Figure ~\ref{fig:tkl} is therefore representative
of the sample statistics.
%----------- new

The $t^*$ distribution is fitted with an exponential function over the range 6$<t^*<$24.8 ns. This corresponds to a time interval $T=\Delta t^*/\tau$\ab0.37.  
With $\sim 8.5 \times 10^6$ events in the fit region we obtain:
$$\tau = (50.87 \pm 0.17)\ \mbox{ns}$$
with a $\chi^2$-value of 58 for 62 degrees 
of freedom (figure \ref{fig:tkl}). 

%--------------------------------------------------------------------------
\section{Systematic uncertainties}
%--------------------------------------------------------------------------
The number of $\kl \to 3\pi^0$ decays at the end of the selection is given by:
\begin{equation}
%\kern-15mm N_{3\pi^0}(l_K)=N_{3 \pi^0}(0)\!\int\!\epsilon_{tot}(l'_K)\!\x\! 
%e^{-l'_K/\lambda'}\!\x\! e^{-(l_K-l'_K)^2/2\sigma^2(l'_K)} dl'_K+N_{bck}(l_K),
\kern-15mm N_{3\pi^0}(l_K)=N_{3 \pi^0}(0)\!\int\!\epsilon_{tot}(l'_K)\!\x\! 
e^{-l'_K/\lambda'}\!\x\! g(l_K-l'_K) dl'_K+N_{bck}(l_K),
\label{eqn:nkl_real}
\end{equation}
where $N_{bck}$ is the residual background at the end of the signal selection, $\epsilon_{tot}(l'_K)$ is the 
signal efficiency ($\epsilon_{\rm tot} = \epsilon_{tag}\x\epsilon_{sel}$) and
$g(l_K-l'_K)$ is the vertex resolution function. 
Finally $\lambda_E = (1/\lambda_L + 1/\lambda_I)^{-1}$
is the effective mean decay 
length taking into account the $\kl$ interactions inside 
the chamber (in the gas mixture and wires) 
and $\lambda_L$ is the mean \kl\ decay length.

Many effects distort the proper time distribution and have been corrected for.
The uncertainty in the corrections is included in
the systematic error on the $K_L$ lifetime.
%-------------------------------------------------------
% Systematic uncertainty due to the tagging efficiency
%-------------------------------------------------------
As noted in Section 3, the tagging efficiency (figure ~\ref{fig:tag}) is well
described by a linear function of $l_K$ 
with a constant term $a = (68.04 \pm 0.03) \%$ 
and a slope $b$=\pt(1.2 \pm 0.2),-5,/cm. 
While the tagging efficiency is easily parametrised, it 
has a significant effect on the overall statistics of the sample.
Therefore, the value of the lifetime is corrected for the effects 
of the $l_K$ dependence of the tagging efficiency using an 
analytical correction: 
$\lambda^* \simeq \lambda (1+(b/a)\lambda)$ \cite{K-note}.
This results in a correction on the lifetime of $ -0.6 \%$ with a 
systematic uncertainty of $\pm 0.1 \%$.

We also vary the threshold of the cluster energy of the
pions from $K_S$ from 40 to 70 MeV in 10 MeV steps.
The slope changes by 0.5\% with a systematic uncertainty of $\pm 0.25\%$.

%-------------------------------------------------------
% Systematic uncertainty due to the selection efficiency
%-------------------------------------------------------
As discussed, the vertex reconstruction 
efficiency has been corrected for its dependence on $l_K$.
We assign a systematic error of $\pm$0.2\% due to the statistical uncertainty on the slope of the data/Monte Carlo efficiency ratios evaluated with $\kl \to \pppco$ events.

We investigate the effects of the cluster energy threshold for photons 
$E_{thr}$, by varying $E_{thr}$ from 10 to 35 MeV in steps of 5 MeV and 
repeating the full analysis.
%-------------------------------new
 The value of this threshold affects dramatically both the 
 background contamination and the relative 
 weights of the samples of different photon-cluster multeplicity. 
 For example, if $E_{thr}$ goes from 20 MeV to 15 MeV, 
 the relative weight of the 
 three photon-cluster sample is reduced by almost a factor 2 
 while the background increases by 20 \%, affecting mainly the
 four photon-cluster sample. Nevertheless, 
%-------------------------------
the fit changes by $\leq\pm0.2\%$ for $15<E_{thr}<35$ MeV,  
which we take as a systematic uncertainty. 
The total systematic uncertainty due to the event selection 
is therefore $\pm 0.3\%$.
%-------------------------------------------------------
% Systematic uncertainty due to the vertex resolution
%-------------------------------------------------------
%Finally the effect of the vertex resolution on the fit result is below
%0.1$\%$, which we neglect.

%------------------new
The effect of the vertex resolution on the fit value has been studied by 
smearing the values sampled from an exponential function with 
the measured $\sigma$'s (as a function of $l_K$)
both in the case of a single-gaussian fit and in 
the case of a double-gaussian fit \cite{K-note}. 
In the second case, the smearing is performed by taking into account the
relative weights of the two gaussians at a given $l_K$.
When the fit is performed to a generated sample of $l_K$ values
without or with a smearing with the known $1 \sigma$
resolution parameters, the lifetime value changes by well under 0.1 \%.
Resolution effects are thus negligible in determining the final value of $\kl$
lifetime.
However, if the resolution parameters $\sigma$ are $\sim 10 \%$ larger
than the measured values, the effect on the 
value of the $\kl$ lifetime is $\sim 0.1\%$. 
The $\sigma$'s are know at 1-2 \% level.
For the systematic uncertainty in the lifetime value 
for vertex resolution effects we assign a 
symmetric 0.1 \% error, which is conservately based 
on the assumption that the resolution parameters can 
be underestimated by as much as 10 \%.
%------------------new

%-------------------------------------------------------
% Systematic uncertainty due to nuclear interactions
%-------------------------------------------------------

$\kl$ interactions with the material inside the chamber 
bias the lifetime measurement since they 
reduce the \kl\ mean path by $(1+\lambda_L/\lambda_I)^{-1} \sim (1-\lambda_L/\lambda_I)$ \cite{K-note}.
The interaction rates for regeneration and $\Lambda$ or $\Sigma$ production are determined from data  \cite{km302}.
The contribution of the regeneration in the DC material is found to be \ab\x3 times lower in data than in 
the MC prediction.
In data the contribution of the total nuclear 
interactions is $\sim 0.33 \%$ to which 
we assign a conservative error of 50\%, (0.33$\pm$0.16)$\%$.
Therefore, the \kl\ lifetime is corrected by +0.33$\%$ with a systematic uncertainty of 0.16\%.

%-------------------------------------------------------
% Systematic uncertainty due to background
%-------------------------------------------------------

The background (table \ref{tab:back2}) is subtracted 
from the proper time distribution using MC simulation.
%----------------new
A residual correction of $+0.2 \%$ due to background subtraction has been added
due to the fact that the background from the $K_S$ regeneration in the drift
chamber material is a factor of three higher in the Monte Carlo than in data
\cite{km302}. 
Therefore, from Table \ref{tab:back2},
the 0.28$\%$ contribution from the $\kl \to  \ks \to \ppo$ 
reaction must be reduced to $0.1\%$ and the global amount of background 
contamination from 1.32 $\%$ to $1.1\%$.
This directly increases the fit value of the lifetime by $+0.2\%$.
%-----------------new

\begin{figure}[hb]
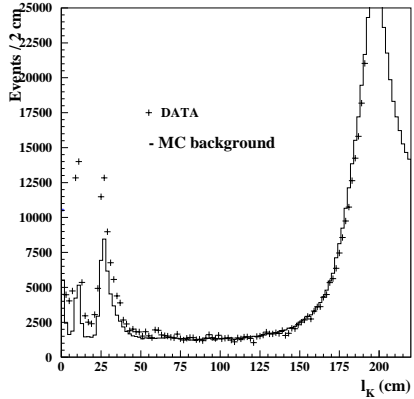

\centering 
\figbox bkg_34;6;
\caption{Decay length distribution for background 
with three and four clusters. Crosses are data, 
solid histogram is Monte Carlo.}
 \label{fig:back34}
\end{figure}

An additional systematic error is due to uncertainties 
in the background scale and $l_K$ dependence in the three 
and four photon-cluster samples.
%---------------- new start
Comparing bin-by-bin data and MC $l_K$ distributions
for background with the combined three and four-cluster event samples 
after MC signal subtraction (figure ~\ref{fig:back34}), 
we find an agreement
at the $\sim 2 \% $ level, with a small linear dependence on $l_K$.
Although the agreement is at the $\sim 2\%$ level, 
we have conservatively taken
the uncertainty in the overall background scale to be $\pm 10 \%$.
The uncertainty in the background scale 
produces a systematic uncertainty in the lifetime value
of $\pm 0.2 \%$.
Correction for the background slope changes 
the fit result by +0.15\% with an uncertainty of $\pm$ 0.06\% 
due to the statistical precision of the slope value.

%--------------- new end

%-----------------------------------------------------------------
% Systematic uncertainty due to the momentum scale and time scale
%-----------------------------------------------------------------
Uncertainties in the DC momentum scale and the absolute EMC 
time scale enter directly in the proper time evaluation $t^* = l_K/(\beta\gamma c)$ and 
give systematic errors respectively of $\pm 0.1\%$ \cite{K-OFFLINE}
and $\pm 7 \x 10^{-4}$ (figure \ref{FIG1}, right).

%-------------------------------------------------------
% Fit stability  with fit limits
%-------------------------------------------------------

The fit stability has been checked by changing the lower 
limit of the time interval used in the fit between 
6 ns and 12 ns and the upper between 21 ns and 28 ns, independently. 
No change in $\tau_L$ is found within its statistical error.
We have also checked the fit stability vs polar angle dividing 
the fiducial volume in two regions containing the same number of events. 
Specifically we chose events with $0.342<|\cos\theta|<0.766$ and 
$|\cos\theta| <0.342$. 
The values of $\tau_L$ from the two zones are consistent to within the statistical accuracy.

%-------------------------------------------------------
% Summary
%-------------------------------------------------------

In table \ref{tab:systematic} we summarize the 
corrections to be applied to the lifetime fit result 
and the corresponding systematic uncertainties. 

\begin{table}[ht]   
\bc
\renewcommand{\arraystretch}{1.1} % enlarge line spacing
\begin{tabular}{|l|c|c|} \hline
  source   &  correction & systematic uncertainty \\ \hline
\vvv tagging efficiency      &  $-0.6 \%$  & $ 0.25 \% $ \\
  acceptance  &  bin by bin  & $ 0.3 \%$ \\
  selection  efficiency   &  bin by bin  & $ 5 \cdot 10^{-5}$ \\
  vertex resolution       &      -       & $ 0.1\%$ \\
  background subtraction        & bin by bin, +0.2\% & $ 0.2\%$ \\
  background shape        & +0.15\%   & $ 0.06\%$ \\
  nuclear interactions    & +0.33\%  & $ 0.16 \%$ \\ 
  momentum scale          &      -       & $ 0.1\%$ \\
  time scale              &      -       & $ 0.07\%$ \\ \hline
  total  & +0.1$\%$ & $0.49\%$ \\ \hline
\end{tabular}
\ec\vglue2mm
\caption{Summary of corrections and systematic uncertainties.
\label{tab:systematic}}
\end{table}

The corrections add to +0.1\% and the central value of the fit is moved accordingly.
The systematic error of $0.49\%$ is at present dominated by the uncertainty 
on the dependence of tagging efficiency with $l_K$ 
and by background subtraction.
The final result is:
\[
  \tau_{\kl} = (50.92 \pm 0.17_{\mbox{stat}} \pm 0.25_{\mbox{syst}}) {\mbox{ ns}} = (50.92 \pm 0.30) {\mbox{ ns}}
\]
This result differs by 1.2 $\sigma$ from the other direct measurement
$\tau_{\kl} = (51.54 \pm 0.44)$ ns \cite{vosburgh:tkl} and by \ab1.8 $\sigma$ from the PDG 2004 fit \cite{pdg2004}, $\tau_{\kl} = (51.8 \pm 0.4)$ ns.

%----------------------
\ack
%----------------------

We thank the DAFNE team for their efforts in maintaining low background running 
conditions and their collaboration during all data-taking. 
We want to thank our technical staff: 
G.F.Fortugno for his dedicated work to ensure an efficient operation of 
the KLOE Computing Center; 
M.Anelli for his continuous support to the gas system and the safety of the
detector; 
A.Balla, M.Gatta, G.Corradi and G.Papalino for the maintenance of the
electronics;
M.Santoni, G.Paoluzzi and R.Rosellini for the general support to the
detector; 
C.Piscitelli for his help during major maintenance periods.
This work was supported in part by DOE grant DE-FG-02-97ER41027; 
by EURODAPHNE, contract FMRX-CT98-0169; 
by the German Federal Ministry of Education and Research (BMBF) contract 06-KA-957; 
by Graduiertenkolleg `H.E. Phys. and Part. Astrophys.' of Deutsche Forschungsgemeinschaft,
Contract No. GK 742; 
by INTAS, contracts 96-624, 99-37; 
and by the EU Integrated Infrastructure
Initiative HadronPhysics Project under contract number
RII3-CT-2004-506078.

%----------------------
%----------------------

\bye